\documentclass[aip,reprint,jcp,floatfix]{revtex4-1}
\usepackage[T1]{fontenc}
\usepackage{graphicx}

\begin{document}

\title{Absorption and luminescence spectroscopy of mass-selected Flavin Adenine Dinucleotide mono-anions}

\author{L.~Giacomozzi}
\affiliation{Department of Physics, Stockholm University, Stockholm, Sweden}

\author{C.~Kj{\ae}r}
\affiliation{Department of Physics and Astronomy, Aarhus University, Aarhus, Denmark}

\author{J.~Langeland~Knudsen}
\affiliation{Department of Physics and Astronomy, Aarhus University, Aarhus, Denmark}

\author{L.~H.~Andersen}
\affiliation{Department of Physics and Astronomy, Aarhus University, Aarhus, Denmark}

\author{S.~Br{\o}ndsted~Nielsen}
\affiliation{Department of Physics and Astronomy, Aarhus University, Aarhus, Denmark}

\author{M.~H.~Stockett}
\affiliation{Department of Physics, Stockholm University, Stockholm, Sweden}

\begin{abstract}
We report the absorption profile of isolated Flavin Adenine Dinucleotide (FAD) mono-anions recorded using Photo-Induced Dissociation action spectroscopy. In this charge state, one of the phosphoric acid groups is deprotonated and the chromophore itself is in its neutral oxidized state. These measurements cover the first four optical transitions of FAD with excitation energies from 2.3 to 6.0~eV (210--550~nm). The $S_0\rightarrow S_2$ transition is strongly blue-shifted relative to aqueous solution, supporting the view that this transition has significant charge-transfer character. The remaining bands are close to their solution-phase positions. This confirms that the large discrepancy between quantum chemical calculations of vertical transition energies and solution-phase band maxima can not be explained by solvent effects. We also report the luminescence spectrum of FAD mono-anions \textit{in vacuo}. The gas-phase Stokes shift for $S_1$ is 3000~cm$^{-1}$, which is considerably larger than any previously reported for other molecular ions and consistent with a significant displacement of the ground and excited state potential energy surfaces. Consideration of vibronic structure is thus essential for simulating the absorption and luminescence spectra of flavins.
\end{abstract}

\maketitle

\section{Introduction}

Flavin Adenine Dinucleotide (FAD) is a ubiquitous redox cofactor serving many key metabolic roles for example as an electron acceptor in the citric acid cycle. FAD is a member of the flavin family which also includes Flavin Mononucleotide (FMN) and riboflavin (RF). These molecules all share the tri-cyclic iso-alloxazine chromophore, whose high reduction potential and multiple redox states give a versatility which lends itself to a wide variety of reactions. In addition, FAD and FMN act as blue light sensors in enzymes and proteins regulating DNA repair \cite{Massey2000}, phototropism and circadian rhythms in plants \cite{Chaves2011} and the perception of magnetic fields by some migratory birds \cite{Solovyov2012,Wiltschko2014}. 

Not unlike other biochromophores such as chlorophyll \cite{Kjaer2016,Wellman2017}, the protein micro-environment may alter the electronic absorption and emission spectrum of flavin cofactors \cite{Udvarhelyi2015}.  For example, the cellular redox equilibrium may favor one or another resting redox state of flavin \cite{Liu2010}, which have rather different optical properties. The redox-specificity of flavin fluorescence has been exploited in autofluorescence imaging applications, where flavins serve as a non-invasive intrinsic biomarker of metabolic activity \cite{Galban2016}. Even for a given redox state, significant differences in the optical spectra and excited state lifetimes of flavins in different proteins have been observed \cite{Kao2008}. In order to quantitatively understand such effects, the intrinsic optical spectra of isolated flavins are useful as a baseline for comparison. Such studies are readily compared to high-level theoretical calculations and eliminate the potentially confounding influence of a solvent. The electronic structure of flavins has been called ``a difficult case for computational chemistry,''\cite{Wu2010} and many authors have bemoaned the dearth of experimental benchmarks \cite{Sikorska2004,Hasegawa2007,Salzmann2008}. To date, only a few optical spectra of isolated flavin-related compounds have been published, including fluorescence and fluorescence excitation spectra of lumiflavin in helium nanodroplets \cite{Vdovin2013}, and action spectra of protonated lumichrome \cite{Guenther2017} (a flavin derivative lacking a N-10 substituent) and anionic FAD \cite{Stockett2017b} \textit{in vacuo}. All of these studies examined only the lowest singlet excited state of the system in question. 


Here, we give the full UV-Vis absorption profile of FAD mono-anions \textit{in vacuo}, covering the first four bright transitions with excitation energies from 2.3 to 6.0~eV (210--550~nm). In addition, we report the luminescence spectrum of FAD mono-anions \textit{in vacuo}. In this charge state, FAD is deprotonated on one of the phosphoric acid groups linking the flavin and adenine moeties (Figure \ref{fig_pid}), and the flavin chromophore is in its neutral, oxidized form \cite{Stockett2017b}. These new experimental results, in conjunction with previously reported solution-phase data, are used to critically evaluate the state of the art in modeling the electronic structure of flavins. 


\section{Experiments}

Absorption profile measurements were performed at two different instruments, the SepI accelerator mass spectrometer complex \cite{Stochkel2011,Wyer2012} and the ELISA electrostatic ion storage ring \cite{Nielsen2001,Andersen2002}, both located at Aarhus University. In both cases, photo-absorption was measured indirectly by Photo-Induced Dissociation (PID) action spectroscopy. Flavin adeine dinucleotide disodium salt hydrate was purchased from Sigma Aldrich and dissolved in methanol. FAD anions were transferred to the gas phase via electrospray ionization and stored in a multipole ion trap which was emptied every 25~ms (SepI) or 40~ms (ELISA). Ion bunches extracted from the trap were accelerated to kinetic energies of 50~keV (SepI) or 22~keV (ELISA) and the ions of interest were selected using a bending magnet according to their mass-to-charge ratio. A high-intensity pulsed laser system (EKSPLA OPO) was used to excite the mass-selected ion bunches \textit{in vacuo}. In action spectroscopy, it is usually assumed that the electronically excited system ultra-rapidly crosses over to a highly vibrationally excited level of the ground electronic state (Internal Conversion), and that this vibrational energy is re-distributed over all internal degrees of freedom in a matter of picoseconds (Internal Vibrational Redistribution). During the $\sim$5~ns irradiation time, the ions may (or may not) absorb multiple photons sequentially, \textit{i.e.} the ion returns to its electronic ground state in between each photon absorption. The deposited energy leads to unimolecular dissociation and/or thermionic electron emission on timescales ranging up to several milliseconds. By monitoring the yield of the photo-products (daughter ions or neutral fragments) as a function of excitation wavelength (corrected for the variation in laser power/photon flux across the spectral range), the so-called action spectrum is constructed. It should be kept in mind that the action spectrum may not perfectly reflect the absorption cross section due to limitations such as sampling time or alternative relaxation channels such as photon or electron emission. Additional experimental details are presented in the Supplementary Material.

\begin{figure}
\includegraphics[width=0.99\columnwidth]{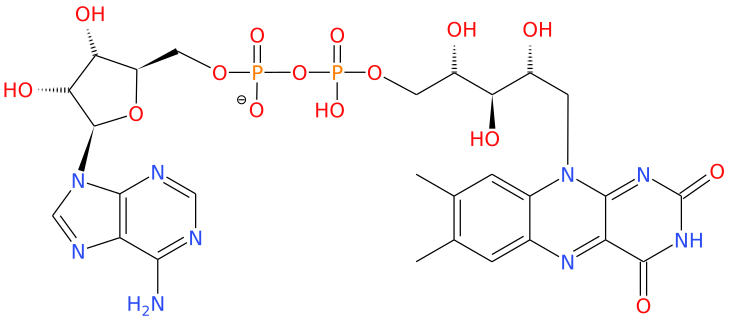}
\includegraphics[width=0.99\columnwidth]{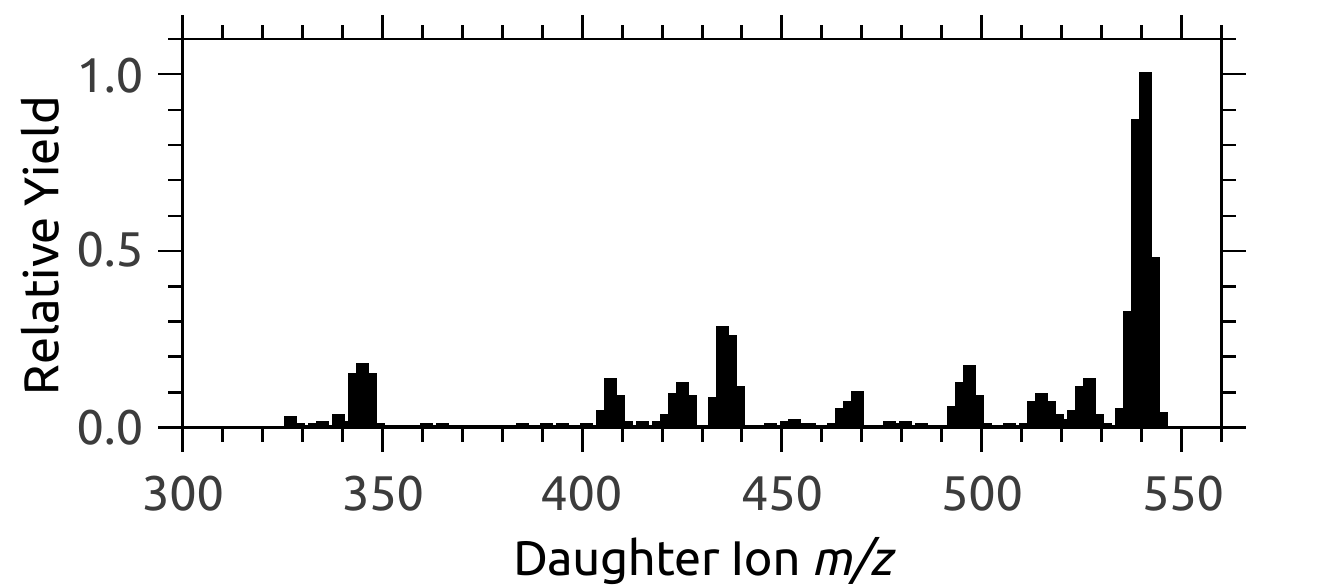}
\caption{Top: Structure of FAD mono-anions. Bottom: Photo-Induced Dissociation mass spectrum of FAD mono-anions (parent ion 784~$m/z$) recorded at SepI with 250~nm (5~eV, 1.2~mJ/pulse) excitation.}
\label{fig_pid}
\end{figure}

At SepI, daughter ions were separated using an electrostatic energy analyzer (mass resolving power $\sim$100) positioned after the laser-ion interaction region and counted with a channeltron detector. Every second ion bunch was irradiated with the laser and the difference in counts between the "laser-on" and "laser-off" injections is the photo-induced signal. The low background rate and high detection efficiency of daughter ions provides superior signal-to-noise than measurement of the depletion of parent ions, particularly for large molecules like FAD with many internal degrees of freedom and correspondingly low dissociation yields. The depletion of the parent FAD mono-anion ion beam measured with 210~nm excitation was 0.8$\pm 0.5\%$. The SepI instrument samples photo-induced dissociation occurring during the $\sim 10$~$\mu$s it takes for the ions to travel from the laser interaction region to the electrostatic analyzer. This limited sampling time could in principle skew the absorption profile towards the blue, an effect known as a kinetic shift. The PID mass spectrum recorded with 250~nm (5~eV) excitation is shown in Figure \ref{fig_pid}. The dominant daughter ion is that with 542~$m/z$. This corresponds to the loss of neutral lumichrome (the flavin rings plus a hydrogen atom), which is the main product of normal photolysis of flavins in solution \cite{Holzer2005}.

\begin{figure}
\includegraphics[width=0.99\columnwidth]{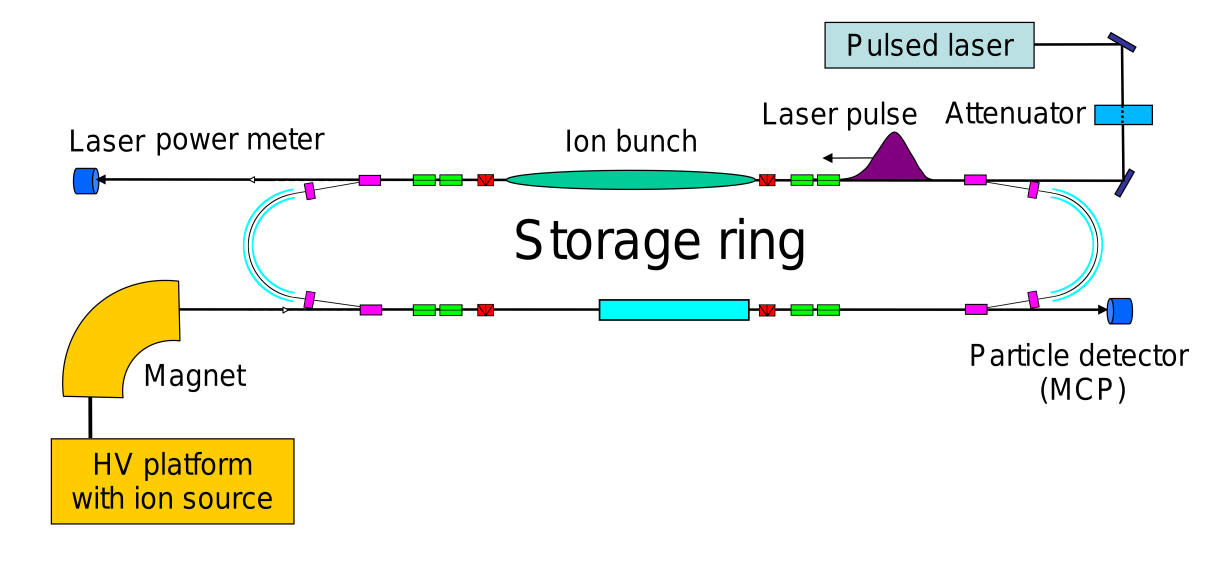}
\caption{ELISA electrostatic ion storage ring. FAD mono-anions circulating in the ring are overlapped with a laser pulse in the upper straight section. Neutral products of dissociation occurring in the lower straight section are detected by the MCP.}
\label{fig_elisa}
\end{figure}

Figure \ref{fig_elisa} shows the ELISA electrostatic ion storage ring. FAD mono-anions circulate around the race-track like ring with a revolution time around 60~$\mu$s. After being stored for 11~ms the laser pulse is overlapped with the ion bunch in the upper straight section. Laser-excited ions may then continue to circulate for several ms before decaying. If they dissociate while in the lower straight section (\textit{i.e.} after at least one half revolution) the neutral fragments will no longer be affected by the electrostatic confinement fields and fly to the microchannel plate (MCP) detector mounted on this section. 

\begin{figure}
\includegraphics[width=0.99\columnwidth]{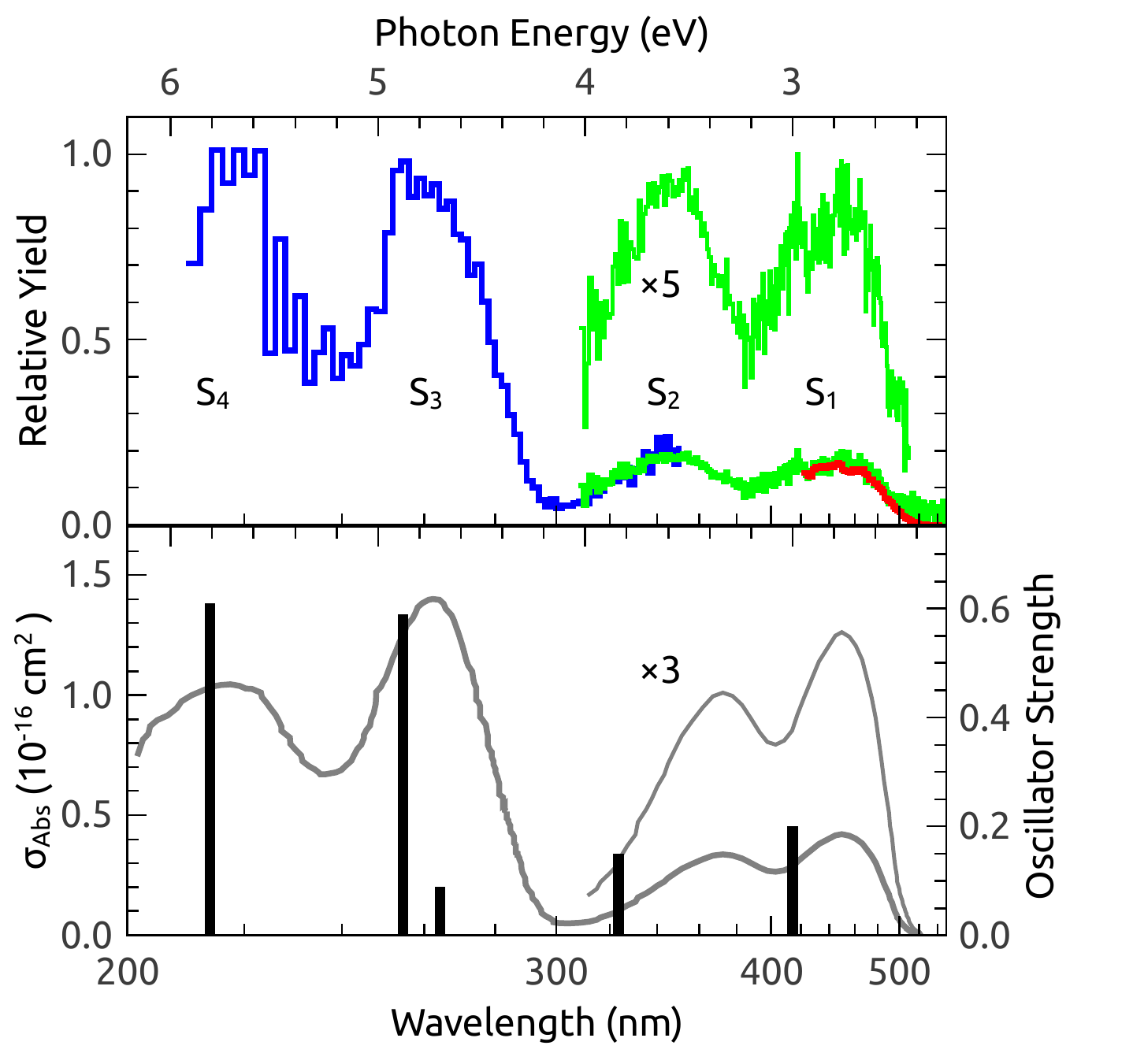}
\caption{Top: PID action spectra for FAD mono-anions. The blue and red curves were recorded using SepI monitoring the yield of the daughter ion with 542~$m/z$. The green curve was recorded using ELISA monitoring the total neutral fragment yield. Bottom: Absorption cross section of FAD in neutral aqueous solution, adapted from Islam \textit{et al.} \cite{Islam2003} The stick spectrum is the consensus of DFT calculations of vertical transition energies for lumiflavin \textit{in vacuo} \cite{Neiss2003,Sikorska2004,Zenichowski2007,Choe2007,Salzmann2008,Vdovin2013,Zanetti-Polzi2017}.}
\label{fig_abs}
\end{figure}

The luminescence spectrum of gas-phase FAD mono-anions was recorded using the LUNA luminescence spectrometer in Aarhus \cite{Stockett2016}. Ions were again produced by electrospray ionization and accumulated in a cylindrical Paul trap. The amplitude and DC offset of the radio frequency trapping voltage applied to the cylinder electrode were set to apply a low-mass cutoff of approximately 600~$m/z$ \textit{i.e.} higher than any of the daughter ions observed in the PID mass spectrum (Figure \ref{fig_pid}). The luminescence signal rate was insufficient to further optimize the mass selection parameters. The trapped ions were excited at 445~nm by an EKSPLA OPO laser system. The laser power was reduced to 50~$\mu$J/pulse to reduce multiple-photon absorption. Luminescence was collected through one of the end caps of the Paul trap, which is made of a wire mesh. An aspheric condenser lens mounted directly behind the mesh collimates the emission which was transmitted through a vacuum window, a 450~nm longpass edge filter (to reduce scattered laser light) and coupled into the entrance slit of an Andor 303i Czerny-Turner imaging spectrograph equipped with a NewtomEM electron multiplying CCD detector array. To correct for scattered laser light and other background sources, the experiment was repeated in alternating sets of 100 cycles with ions in the trap followed by 100 cycles with no ions (the trapping voltage was switched off). The difference between the "ions-on" and "ions-off" acquisitions is the luminescence signal.

\section{Absorption Results and Discussion}

In the upper panel of Figure \ref{fig_abs}, PID action spectra of FAD mono-anions, recorded in three overlapping spectral regions, are shown. SepI was used for the wavelength ranges 210--350~nm and 420--550~nm, monitoring the yield of the daughter ion with 542~$m/z$ (lumichrome loss). ELISA was used in the range 309--550~nm, monitoring the total neutral fragment yield. The SepI data from 420 to 550~nm was reported previously \cite{Stockett2017b}, and is reproduced here to show consistency between the the two measurement techniques. The excellent agreement in the low-energy range between the two datasets, which sample different dissociation timescales, indicates that our results are not strongly affected by any kinetic shift. The lower panel of Figure \ref{fig_abs} shows the absorption cross section of FAD in neutral aqueous solution adapted from Islam \textit{et al.} \cite{Islam2003} With the exception of the $S_0\rightarrow S_2$ transition, which is red-shifted by 0.22~eV (23~nm) in solution, the band maxima are identical within experimental accuracies. Hints of vibronic structure are present in the gas-phase spectrum (upper panel), such as a minor peak at 411~nm, but the bands are generally broad and featureless, consistent with the solution-phase measurements (lower panel). 

%
%
%
%
%
%

\begin{table}
\begin{tabular}{c|cccccc}

$S_0\rightarrow$ & FAD$^a$ & FMN$^a$ & LF$^b$ \cite{Neiss2003,Sikorska2004,Zenichowski2007,Choe2007,Salzmann2008,Vdovin2013,Zanetti-Polzi2017} & RF \cite{Sikorska2005} & FMN \cite{Kammler2012} & FAD$^c$ \cite{Islam2003} \\ 

\hline 

$S_1$ & 2.74 & - & 3.00 & 3.08 & 2.56/2.89 & 2.74 \\ 

$S_2$ & 3.59 & - & 3.84 & 3.68 & 3.26/3.3 & 3.37 \\ 

$S_{3a}$ & - &  - & 4.70 & 4.83 & - & - \\

$S_{3b}$ & 4.75 & 4.69 & 4.88 & 4.89 & - & 4.73 \\ 

$S_4$ & 5.8 & 5.6 & 5.81 & 5.79 & - & 5.78 \\ 

\end{tabular} 
\caption{Band maxima of action spectrum of FAD and FMN mono-anions \textit{in vacuo} compared to electronic structure calculations of vertical transition energies for various flavins \textit{in vacuo} and absorption band maxima of FAD in neutral aqueous solution. All in eV.\\ $^a$Present work, experiment \textit{in vacuo}, uncertainties implied by the number of significant digits\\ $^b$Consensus of TD-DFT values from various authors, see Supplementary Material\\ $^c$Experiment in aqueous solution}
\label{tab_gas}
\end{table}

In Table \ref{tab_gas}, our experimental results are compared with a survey of previously published calculations of gas-phase transition energies for various flavins. Most electronic structure calculations of flavins focus on lumiflavin (LF), the smallest subunit which shares the essential photophysical properties of the larger flavin cofactors. In LF, the ribityl sidechain at the N-10 position is replaced by a methyl group, which simplifies calculations. Most of the calculations of LF have been performed using Time Dependent Density Functional Theory (TD-DFT) methods \cite{Neiss2003,Sikorska2004,Zenichowski2007,Choe2007,Salzmann2008,Vdovin2013,Zanetti-Polzi2017} and show a high degree of consistency (see Supplementary Material for a complete tabulation). Indeed, the variation in transition energies amongst the various TD-DFT calculations is less than 5$\%$. For the sake of comparison, these transition energies (and their calculated transition $f-$values) have been simply averaged to give a ``consensus'' spectrum, which is presented in the lower panel of Figure \ref{fig_abs} \footnote{The SAC-CI calculations by Hasegawa \textit{et al.} \cite{Hasegawa2007} for LF, while agreeing with the qualitative description of the orbitals, yield significantly different transition energies, particularly for $S_0\rightarrow S_1$ which is 0.5~eV below the other calculations for LF and 0.28~eV below the present experimental value for FAD. This value is not included in the consensus spectrum in Figure \ref{fig_abs}.}. The consensus vertical transition energies (given in Table \ref{tab_gas}) overestimate the present experimental band maxima by about 0.3~eV for the first two bands and 0.1~eV for the UV bands. 

All calculations agree that the bright transitions are due to $\pi\rightarrow\pi^*$ transitions. While not all authors provide detailed assignments, generally the $S_0\rightarrow S_1$ is considered to be the HOMO$\rightarrow$LUMO transition and $S_0\rightarrow S_2$ a transition to the LUMO from a lower orbital (usually HOMO-1) which is localized to the aromatic benzene-like ring of the flavin chromophore \cite{Hasegawa2007,Zenichowski2007,Salzmann2008}. This localization gives this transition a significant degree of charge transfer character \cite{Hasegawa2007}, which is widely thought to be responsible for the solvatochromic behavior of $S_0\rightarrow S_2$. The HOMO and LUMO, in contrast, are both spread across the entire chromophore \cite{Hasegawa2007,Zenichowski2007,Salzmann2008} and this transition shows little solvatochromism in theory \cite{Zenichowski2007} or experiment \cite{Sikorska2004,Sikorska2005,Zirak2009}. The present experimental results qualitatively support this view, with a large blue-shift (0.22~eV) upon desolvation for $S_0\rightarrow S_2$, but no such shift for $S_0\rightarrow S_1$.

Several authors have investigated whether solvent effects can explain the large deviation between calculated vertical excitation energies and experimentally measured absorption band maxima in solution \cite{Hasegawa2007,Zenichowski2007,Wu2010,Zanetti-Polzi2017}. The present results, however, show that such effects are small. As has been pointed out earlier \cite{Salzmann2009a}, the absorption spectra of flavins are hardly affected by the solvent environment \cite{Koziol1965,Koziol1966,Sikorska2004,Weigel2008,Zirak2009}, except of course for the $S_0\rightarrow S_2$ transition. Moreover, calculations \cite{Hasegawa2007} and measurements \cite{Stanley1999} find only a small increase in the permanent dipole moment of the flavin chromophore upon excitation to $S_1$. Solvatochromism measurements \cite{Zirak2009} actually imply a slight \textit{decrease} in the dipole moment upon excitation, but with a large uncertainty. There is thus no reason to expect large solvent effects for $S_0\rightarrow S_1$, and indeed none are found in most calculations \cite{Hasegawa2007,Zenichowski2007}, or from the present gas-phase experiments. 

Setting aside solvent effects, the vibronic structure of flavins must seriously be taken into account. As the density of vibrational levels in an electronically excited state increases strongly with energy, the absorption band maximum is usually observed to the blue of the 0-0 transition energy. The band maximum often roughly coincides with the vertical transition energy calculated in TD-DFT from the ground state equilibrium geometry, although clearly not in the case of flavins. Full calculations of broadened vibronic excitation spectra reported for flavins \cite{Salzmann2009a,Klaumuenzer2010,Goetze2013,Karasulu2014} come closer to reproducing the profile and position of the present gas-phase results than ``simple'' TD-DFT. There remains some discrepancy in the 0-0 energy (0.05~eV \cite{Klaumuenzer2010} to 0.5~eV \cite{Karasulu2014}, depending on the calculation) compared to that measured in He droplets for lumiflavin \cite{Vdovin2013}, which is presumed to be due to limitations in the chosen functionals. These methods have struggled to include micro-solvation effects, and are rarely attempted for excited states higher than $S_1$. We hope the present contribution will serve as a benchmark for further refining these methods, as it is clear that careful consideration of vibronic activity is essential in modeling absorption by flavins.


Solvent effects are evidently important for modeling the $S_0\rightarrow S_2$ transition. The $S_2$ absorption band maximum of riboflavin varies from 332~nm (3.75~eV) in apolar dioxane to 367~nm (3.38~eV) in water \cite{Koziol1965,Sikorska2005}. The present value for the gas phase is 346~nm (3.59~eV). Calculations \cite{Hasegawa2007} and measurements \cite{Stanley1999} find that the permanent dipole moment of the $S_2$ state is significantly higher than that of $S_0$, implying bulk polarization effects may be important. In addition, significant differences between polar aprotic solvents such as DMSO and polar protic solvents like water suggest that hydrogen bonding plays a role as well \cite{Zirak2009}. Calculations including both of these effects do well at reproducing the magnitude of the solvent shift of $S_0\rightarrow S_2$ \cite{Hasegawa2007,Salzmann2008}.

A few TD-DFT calculations have been performed on more complex flavins including the ribityl sidechain. The addition of the sidechain appears to have little influence on the HOMO and LUMO orbitals \cite{Sikorska2005,Wolf2008,Klaumuenzer2010,Wu2010}. However, the sidechain may participate in other orbitals, notably including those involved in the  $S_0\rightarrow S_2$ transition \cite{Wu2010,Kammler2012}. This could affect the degree of charge transfer character in these transitions. Sikorska and coworkers reported theoretical spectra for both LF \cite{Sikorska2004} and RF \cite{Sikorska2005} and found $S_0\rightarrow S_2$ to be nearly 0.2~eV lower for RF, while the other transition energies agreed with the consensus for LF. 

\begin{figure}
\includegraphics[width=0.99\columnwidth]{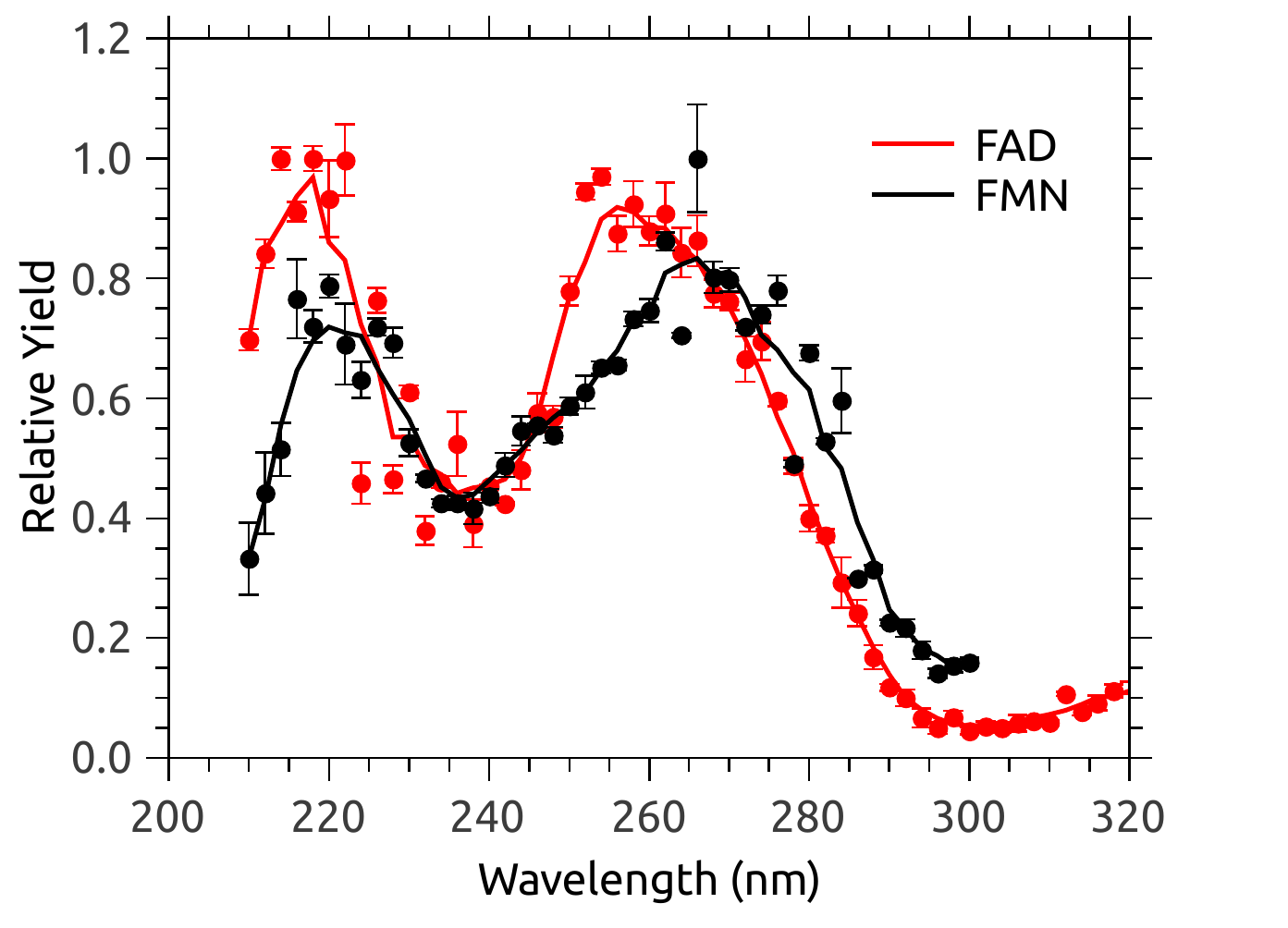}
\caption{Comparision of action spectrum of FAD and FMN mono-anions recorded at SepI. The spectrum for FAD was recorded monitoring the photo-induced yield of daughter ions with 542~$m/z$ (lumichrome loss, same data as main text). For FMN, the yield of 169~$m/z$ (loss of formylmethylflavin) was monitored.}
\label{fig_fmn}
\end{figure}

Interpretation of the UV portion of the absorption profile is complicated by the presence of the adenine moiety in FAD and the relative lack of calculations and solution-phase data in this spectral region. Most calculations find the $S_0\rightarrow S_3$ band of the flavin chromophore to be composed of two transitions (labeled $S_{3a}$ and $S_{3b}$ in Table \ref{tab_gas}), which are not resolved in the present experiment. Adenine absorbs at wavelengths similar to those of flavins, with band maxima near 250 and 200~nm in the gas phase \cite{Li1987}, but with lower absorption cross sections (in solution) \cite{Islam2003}. We are aware of no modern quantum chemical calculations of the full FAD system. To add another point of comparison, we recorded a PID action spectrum of FMN (which lacks the adenine part) mono-anions at SepI. Figure \ref{fig_fmn} shows the PID action spectra of FAD and FMN mono-anions recored at SepI. The FAD spectrum is the same data presented in Figure \ref{fig_abs}, the solid line is a 5-point moving average. The action spectrum for FMN (parent mass 455~$m/z$) was recorded monitoring the yield of daughter ions with 169~$m/z$, the most prominent peak in the PID mass spectrum (not shown). This corresponds to the loss of formylmethylflavin, and was determined to be a 1-photon process. The error bars are the standard deviation of 6 individual scans. Comparing the action spectra, the presence of adenine leads to an apparent blue-shift in the $S_0\rightarrow S_3$ and $S_0\rightarrow S_4$ transitions of about 0.05 and 0.2~eV, respectively. The positions of the band maxima are given in Table \ref{tab_gas}. These  results imply a larger discrepancy between theory and experiment for FMN than for FAD. It should be kept in mind, however, that the dissociation rates and competition with other channels like thermionic electron emission could be different for FAD and FMN, potentially skewing the absorption profiles. It is also not obvious whether the differences in the action spectra are due to absorption by adenine, or by perturbation of the electronic structure of the flavin chromophore. In relation to the this, we note that the present measurement of the $S_0\rightarrow S_1$ band maximum of FAD is slightly red-shifted (by $\sim 0.05$~eV) with respect to most solution-phase measurements of other flavins that lack the adenine moiety \cite{Sikorska2004,Sikorska2005,Zirak2009}. This shift (between FAD and flavins without adenine) is also found in solution-phase measurements \cite{Islam2003,Hussain2016}. As adenine does not absorb in this spectral region, this shift may point to a small stabilizing effect of the adenine moiety on the ground state of the flavin chromophore in FAD.

\section{Luminescence Results and Discussion}

 \begin{figure}
\includegraphics[width=0.99\columnwidth]{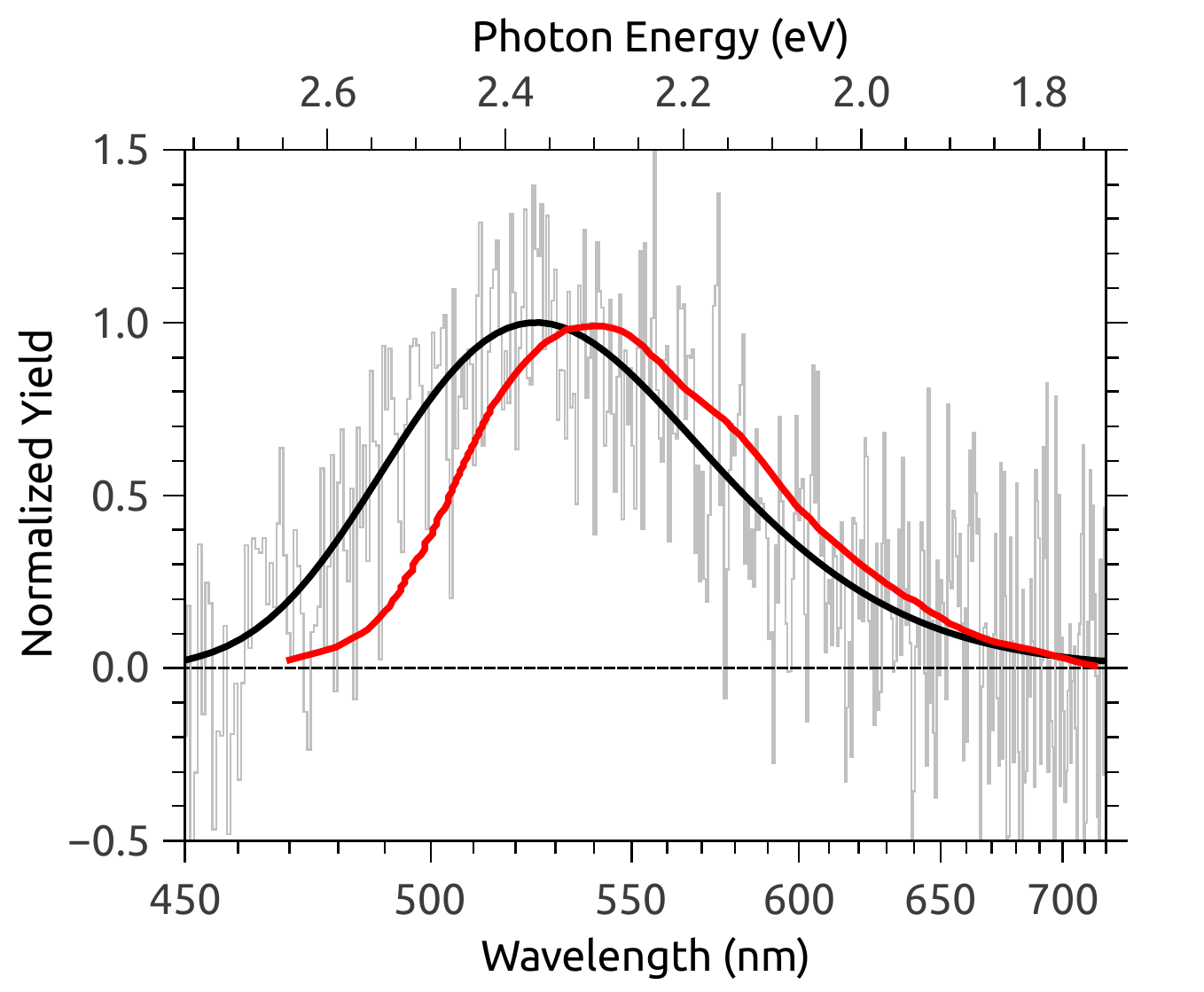}
\caption{Fluorescence spectrum of FAD mono-anions \textit{in vacuo} (black, excitation wavelength 445~nm) and in aqueous solution, adapted from Islam \textit{et al.} \cite{Islam2003} (red, excitation wavelength 428~nm).}
\label{fig_lumi}
\end{figure}

Figure \ref{fig_lumi} shows the luminescence spectrum of FAD mono-anions \textit{in vacuo} excited at 445~nm, as well as that in aqueous solution (adapted from Islam \textit{et al.} \cite{Islam2003}). Both spectra represent $S_1\rightarrow S_0$ fluorescence. To our knowledge, Figure \ref{fig_lumi} is the first reported gas-phase fluorescence spectrum of a completely bare (tag-free), naturally occurring biomolecular ion in its physiological charge state. The light grey line is the raw data from the CCD and the solid black line is a fit to the functional form 

\begin{equation}
y=A\times exp[-exp(-(x-x_0)/w)-(x-x_0)/w],
\end{equation}

\noindent as recommended by Greisch \textit{et al.} \cite{Greisch2014} as an empirical tool for characterizing asymmetric luminescence bands. The fit gives the position of the band maximum $x_0=525\pm 2$~nm (2.37~eV). The position of the fluorescence maximum in water is 541~nm \cite{Islam2003}, which is consistent with other flavins in water \cite{Zirak2009,Weigel2008}. In contrast to $S_0\rightarrow S_1$ absorption band maximum, the fluorescence band maximum of flavins varies significantly with solvent polarity, varying from 542~nm in water to 509.5~nm in benzene for riboflavin \cite{Zirak2009}. This suggests that the fluorescence solvatochromism is due to stabilization of the more polarizable excited state by solvent dipole rearrangement. The gas phase Stokes shift of 3000~cm$^{-1}$ (0.37~eV) is consistent with that of riboflavin in the least polar solvents such as chloroform \cite{Zirak2009}. In polar, protic solvents, Stokes shifts close to 4000~cm$^{-1}$ are observed \cite{Zirak2009,Weigel2008}. We note that the emission maximum and Stokes shift for riboflavin in DMSO are more similar to those in apolar solvents than polar protic ones \cite{Zirak2009}, again suggesting that micro-solvation effects (\textit{e.g.} H-bonding) play some role. 

Notably, the gas-phase Stokes shift for FAD is significantly higher than that measured for other complex molecular ions such as xanthene \cite{McQueen2010,Forbes2011,Kjaer2017} and phenoxazine \cite{Stockett2016b} laser dyes. Gas-phase stokes shifts for rhodamine dyes, for example, have been found to range from 900~cm$^{-1}$ \cite{Wellman2015} to less than 500~cm$^{-1}$ \cite{Forbes2011}. The analysis of Klaum\"{u}nzer \textit{et al.} \cite{Klaumuenzer2010} indicates that several stretching modes of the iso-alloxazine chromophore with frequencies in the range 1400-1600~cm$^{-1}$ dominate the vibronic spectra of flavins and predicts a Stokes shift of 3400~cm$^{-1}$. As has been observed for other complex molecules \cite{Stockett2016b}, these vibrational frequencies correspond to about half the value of the gas-phase Stokes shift. This helps us understand the discrepancy between calculated vertical transition energies and observed absorption band maxima, the correspondence between which relies on the assumption of high vibrational excitation upon absorption such that the vibrational wavefunctions peak at the classical turning points \cite{Davidson1998}. Although the ground and excited state structures of flavins may be significantly displaced from each other, the difference in energy between the vertical and adiabatic (0-0) transition is covered by only a few quanta of the most strongly coupled vibrational modes and thus fails to meet this criterion.

The fluorescence signal detected from FAD mono-anions is very weak. Determination of absolute fluorescence quantum yields of trapped ions is experimentally challenging as key parameters such as the number of ions in the trap and the overlap between the laser beam and the ion cloud are difficult to measure precisely. Instead, the ``brightness'' (the total integrated fluorescence signal per laser shot) is often used to compare the luminescence from different ions recorded under similar experimental conditions \cite{Yao2013,Stockett2017a}. The brightness of FAD mono-anions is at least an order of magnitude lower than that of resorufin, an anionic xanthene dye \cite{Kjaer2017}. If we assume that the fluorescence quantum yield of gas-phase resorufin is the same as in aqueous solution (0.75 \cite{Bueno2002}), we can roughly estimate the gas-phase quantum yield of FAD to be about 0.1 (for additional details, see Supplementary Material). 

While the quantum yields of riboflavin and FMN are reasonably high (around 0.26 \cite{Drossler2002,Islam2003,Sikorska2005,Zirak2009}) in neutral solutions, FAD is thought to exist in a stacked conformation where electron transfer from the adenine moiety quenches the flavin excited state \cite{Berg2002,Li2009}. This leads to a reduced fluorescence quantum yield of 0.033 \cite{Islam2003}. At reduced pH, un-stacked conformation becomes dominant and the quantum yield rises to 0.13 \cite{Islam2003}. 

While our estimate of the gas-phase quantum yield is too crude to distinguish between stacked and non-stacked conformations, it is interesting to note that the quantum yield does not appear to be significantly lower than in solution. In contrast, \textit{no} fluorescence was detected for FAD di-anions or, more remarkably, from FMN anions, using the LUNA instrument. Using the same brightness comparison with resorufin, we can estimate an upper limit on the gas-phase quantum yield of FMN of 0.04, much less than in solution. This suggests that these ions may decay through some non-radiative channel (\textit{e.g.} electron detachment or inter-system crossing) that is not competitive for FAD mono-anions. Changes in fluorescence quantum yield upon desolvation have been reported for other molecules \cite{McQueen2010,Sciuto2015} and may be a more sensitive indicator of changes in photophysics than transition energies. 

\section{Conclusion}

We have reported the photo-induced dissociation action spectrum and the luminescence spectrum of FAD mono-anions \textit{in vacuo}. These results confirm that the vertical transition energies calculated using various electronic structure methods overestimate the intrinsic absorption band maxima. Neglect of vibronic structure, rather than solvent effects, is the cause of this discrepancy. Bulk polarization and micro-solvation effects appear to be important only for the $S_0\rightarrow S_2$ transition, in agreement with calculations which indicate that this transition has significant charge-transfer character. Luminescence emission occurs at a rather high Stokes shift compared to previously reported studies of complex molecular ions \textit{in vacuo} in what again appears to be a vibronic effect. No emission was seen from two other flavin anions. The observed micro-environmental sensitivity of fluorescence portends a prominent role for gas-phase luminescence spectroscopy studies in unraveling the intrinsic photophysics of complex biomolecules.

\section*{Supplementary Material}

The Supplementary Material includes additional experimental details, a tabulation of previously published calculations of transition energies of lumiflavin, and a description of our approach to estimating the gas-phase quantum yield of FAD.

\section*{Acknowledgements}

This work was supported by the Swedish Research Council (grant numbers 2016-03675 and 2016-04181) and the Danish Council for Independent Research (4181-00048B). LG thanks H.~Zettergren of Stockholm University.



%

\end{document}